\documentclass[lettersize,conference]{IEEEtran}
\usepackage{bm}
\usepackage{cite}
\usepackage[pdftex]{graphicx}
\usepackage{multirow}
\usepackage{subfigure}
\usepackage{caption}
\usepackage{multicol}
\usepackage{comment}
\usepackage{color}
\newsavebox{\bigimage}
\usepackage{soul}
\usepackage[dvipsnames]{xcolor}
\usepackage{amsmath}
\usepackage{amssymb}
\usepackage{mathptmx}
\usepackage[left=0.68in,right=0.673in,top=0.76in,bottom=1.007in]{geometry}
\ifCLASSINFOpdf
\else
\fi
\graphicspath{{Figures/pdf/}}
\newcommand{\degree}{\ensuremath{^{\circ}}}
\title{Antennas in Walls: Performance Analysis of Microstrip Patch Antennas Designed for\\ Internet of Paint (IoP)}
\author{\IEEEauthorblockN{L. T. Wedage\IEEEauthorrefmark{1}, Mehmet C. Vuran\IEEEauthorrefmark{2}, Bernard Butler\IEEEauthorrefmark{1}, Christos Argyropoulos\IEEEauthorrefmark{3} and Sasitharan Balasubramaniam\IEEEauthorrefmark{2}}
\IEEEauthorblockA{\IEEEauthorrefmark{1}Walton Institute, South East Technological University, Ireland, 
thakshila.wedage@waltoninstitute.ie, bbutler@ieee.org}
\IEEEauthorblockA{\IEEEauthorrefmark{2}School of Computing, University of Nebraska-Lincoln, USA, \{mcv,sasi\}@unl.edu}
\IEEEauthorblockA{\IEEEauthorrefmark{3}Pennsylvania State University, USA, cfa5361@psu.edu}}

\begin{document}
\maketitle

\begin{abstract}
This study presents a simulated transceiver with a microstrip patch antenna (MPA) designed to resonate at $150$ GHz and embedded in paint. The in-paint MPA (IP-MPA) is designed for the Internet of Paint (IoP) paradigm, which envisions seamless device communication through a paint layer on walls. This study introduces a comprehensive channel model for transceivers in paint at arbitrary depths and IP-MPA orientations. The best antenna orientations are analyzed for IoP channel performance. Extensive simulations indicate that the lateral waves, which propagate along the air-paint interface, exhibit the lowest loss, making this path the most reliable for communication between transceivers in paint. Further, the maximum received power for each propagation path, with the exception of the direct path, depends on depth. The findings suggest that the proposed network of IP-MPA-enabled transceivers for IoP has the potential to transform conventional walls into an integrated high-speed wireless communication and sensing infrastructure.
\end{abstract}


% \fi
\IEEEpeerreviewmaketitle

\section{Introduction}
\label{sec:Intro}

Terahertz (THz) communication operates between microwave and infrared frequencies, providing high bandwidth capacity and data transmission rates, enabling next-generation wireless networks and ultra-high-speed communication systems. The extremely short wavelengths in the THz band allow the development of microscale communication and sensing devices, including advanced patch antennas at millimeter scales. These devices can be embedded within ultra-thin, functionalized surfaces such as conductive paint layers, enabling seamless and highly integrated connectivity across various devices~\cite{Thakshila_IoP}. Combining THz frequencies with miniaturized antennas can also expand functionalities beyond communication by supporting high-resolution sensing.

The Internet of Paint (IoP) concept is introduced in \cite{Thakshila_IoP}, where IoP consists of embedded nano-devices that communicate through the paint in the sub-THz frequency spectrum. IoP can potentially transform a paint layer into a communication surface, providing unprecedented coverage and wall-based connectivity. The IoP channel model and its capacity are analyzed in~\cite{Thakshila_IoP}, but the performance of antennas embedded in a dielectric paint medium, which is the focus of this paper, has not been analyzed. Transceivers for IoP have unique properties because the paint surrounds the antenna, and so the paint can function as both a superstrate and protective cover for the antenna. In \cite{khan2020high}, a high-performance circular patch antenna is developed for THz band applications. These antennas use graphene as the MPA material, with a thin layer of Teflon and glass serving as the superstrate. The antenna operates at $7$ THz and demonstrates high efficiency with a gain of $7.286$ dBi and $7.392$ dBi when Teflon and Glass used as the superstrate, respectively. In \cite{younssi2013study}, a rectangular microstrip patch antenna is presented that operates within the $0.6$ to $0.8$ THz range with an RT/Duroid $6006$ superstrate. They observed a matching bandwidth of $22.47$\% and achieved a maximum radiation gain of $10.43$ dBi at $0.6929$ THz. In \cite{bahl1982design}, a microstrip antenna was proposed, utilizing polystyrene, ice, and beryllium oxide dielectric layers as superstrates. Using these superstrates, the resonant frequency changes by $5.8$\%, $7.8$\%, and $16$\% at $10$ GHz for infinitely thick dielectric covers of polystyrene, ice, and beryllium oxide, respectively. In \cite{Concrete2016analysis}, a compact patch antenna is designed to operate at a frequency of $900$ MHz and embedded in concrete for wireless monitoring applications. Results demonstrate the antenna's effective performance across a wide range of permittivity values for concrete, with minimal change in return loss at the operating frequency. Despite numerous studies emphasizing communication through miniaturized embedded transceivers, a noticeable gap remains in research specifically addressing the design of embedded antennas within the paint. This study aims to fill that void by exploring the performance of micro-scale antennas proposed for IoP.

\section{IoP Antenna Design}
\label{sec:IoP_Antenna_Design}

In this section, we explore the design of a passive in-paint microstrip patch antenna (IP-MPA) resonant at a frequency of $150$\,GHz using the parameters highlighted in Table~\ref{tab:parameters_designing_MPA}. The design of the IP-MPA for IoP is unique due to its integration into the paint medium, which acts as a superstrate. For this study, we consider a thick paint layer ($h_p = 5$ mm) to embed the IP-MPA. When selecting the frequency for IoP, we also consider the thickness of the paint layer and the proportional relationship between the wavelength and the size of the antenna. As a result, we determine that the sub-THz frequency spectrum is suitable for IoP. Thus, we consider a $150$\,GHz resonant frequency for the IP-MPA in this study, while the developed model can capture other resonant frequencies in the THz spectrum with different antenna dimensions. The subject of powering passive IP-MPAs is not the primary focus of this study. Nevertheless, we have discussed the associated challenges related to the powering of IP-MPAs in \cite{Wedage_IoP_challenges}. Note that the resonant frequency of an MPA can vary based on the substrate thickness and dielectric properties. Also, when simulating the IP-MPA, the (THz) wave speed in the superstrate (paint) medium as a factor of the refractive index (i.e., $c_p=c/n_p$, where $c$ is the speed of the wave in free space, $c_p$ is the speed of the wave in paint, and $n_p$ is the refractive index of the paint) should be considered. Further, we utilize \textit{foam} as the substrate for the IP-MPA, a lightweight, porous material with a relatively low dielectric constant ($\epsilon_s$) of $1.03$. 

\begin{table}[t!]
    \centering
    \begin{tabular}{|l|c|}
    \hline
        Parameter & Description\\
    \hline
       Resonant frequency in paint ($f_r$)  & $150$ GHz\\
       Substrate  & Foam \\
       Substrate Dielectric constant ($\epsilon_s$)  & $1.03$ \\
       Substrate Thickness ($h_s$)  & $10~\mu$m\\
       Superstrate  & Titanium White Paint \\
       Superstrate Dielectric constant ($\epsilon_p$)  & $4.5369$ \\
       \hline
    \end{tabular}
    \caption{\label{tab:parameters_designing_MPA} Parameters of the microstrip patch antenna.}
\end{table}

A practical patch antenna operates around $50\,\Omega$, and the patch width ($W_p$) is usually much larger than the substrate thickness ($h_s$). Thus, for a patch antenna with $W_p/h_s > 1$, the effective dielectric constant without the superstrate is \cite{saidulu2013characteristics}:
\begin{equation}
    \epsilon_{eff} = \dfrac{\epsilon_s+1}{2} + \dfrac{\epsilon_s-1}{2}{\left( 1+\dfrac{12h_s}{W_p}\right)^{-\frac{1}{2}}}.
\end{equation}

When the MPA is embedded inside paint, which functions as a superstrate for the MPA, the effective dielectric constant is increased and can be expressed as \cite{Concrete2016analysis}:  
\begin{equation}
    \epsilon_{eff,p} = \dfrac{\epsilon_s+\epsilon_p}{2} + \dfrac{\epsilon_s-\epsilon_p}{2}{\left[ 1+\dfrac{12h_s}{W_p}\right]^{-\frac{1}{2}}},
\end{equation}
where, $\epsilon_p$ is the dielectric constant of the paint.

The microstrip antenna's patch length ($L_p$) appears larger electrically due to fringing effects. Thus, its dimensions extend at each end along its length \cite{saidulu2013characteristics}, and can be expressed as:
\begin{equation}
    L_p = L_{eff} - 2 \left\{ 0.412h_s \dfrac{\left( \epsilon_{eff,p} + 0.3 \right)\left( W_p/h_s + 0.264 \right)}{\left( \epsilon_{eff,p} - 0.258 \right)\left(W_p/h_s + 0.8 \right)} \right\},
\end{equation}
%\commentBB{Where do constants like 0.412 come from?\commentTW{refer to \cite{saidulu2013characteristics}}}
where,
\begin{equation}
    W_p = \dfrac{c_p}{2f_r}\sqrt{\dfrac{2}{\epsilon_s+1}}.
\end{equation}

The effective length of the patch $L_{eff}$ is:
\begin{equation}
\label{resonant_f_in_air}
    L_{eff} = c_p/\left(2f_r \sqrt{\epsilon_{eff,p}}\right).
\end{equation}

The feed line length, $L_f=2 W_f$, where $W_f$ is the feed line width and can be expressed as \cite{qasem2020simulation}:
\begin{subequations}
\begin{align}
    W_f & = \dfrac{2h_s}{\pi}\left[ B_1 -1 - \ln{(2B_1-1)} + B_2 \right],\\
B_1 & = \dfrac{377\pi}{2Z_{in}\sqrt{\epsilon_s}}, \\
B_2 & = \dfrac{\epsilon_s-1}{2\epsilon_s}\left[ \ln{(B_1-1)+0.39-\dfrac{0.61}{\epsilon_s}} \right],
\end{align}
\end{subequations}
where $Z_{in}$ is the input impedance, assumed to be $50~\Omega$. Finally, the substrate width ($W_s$) and the length ($L_s$) are
\begin{equation}
    W_s   = 2 \times W_p, \hspace{0.2cm}
    L_s   = 2 \times L_p.
\end{equation}

\begin{table}
    \centering
    \begin{tabular}{|c|l|}
    \hline
     Dimension & Size\\
     \hline
       $W_p$  & $446$ $\mu m$ \\
       $L_{p}$  & $412$ $\mu m$\\
       $W_f$  & $48$ $\mu m$\\
       $L_f$  & $97$ $\mu m$\\
       $W_s$  & $931$ $\mu m$\\
       $L_s$  & $824$ $\mu m$\\
         \hline
    \end{tabular}
    \caption{\label{tab:MPA_dimensions}In-paint microstrip patch antenna dimensions.}
\end{table}

\begin{figure*}[t!]
    \centering
    \subfigure[]{\includegraphics[width=0.26\linewidth]{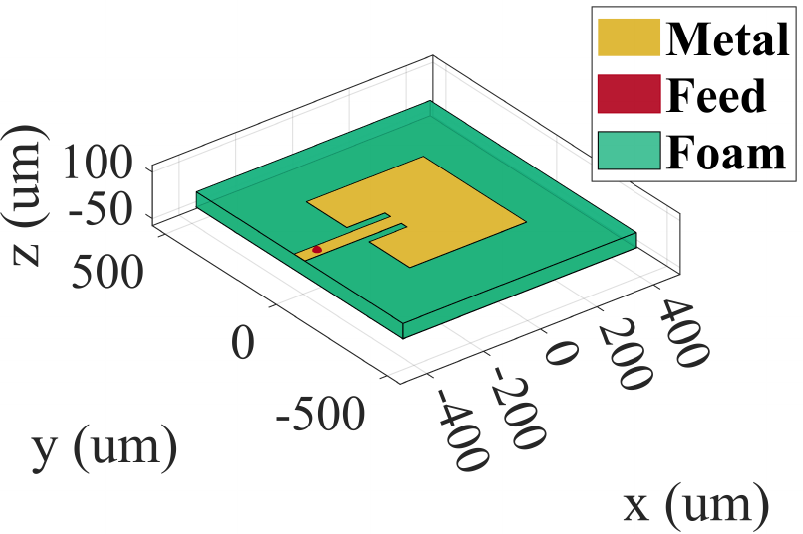}\label{fig:MPA-design}} 
    \subfigure[]{\includegraphics[width=0.22\linewidth]{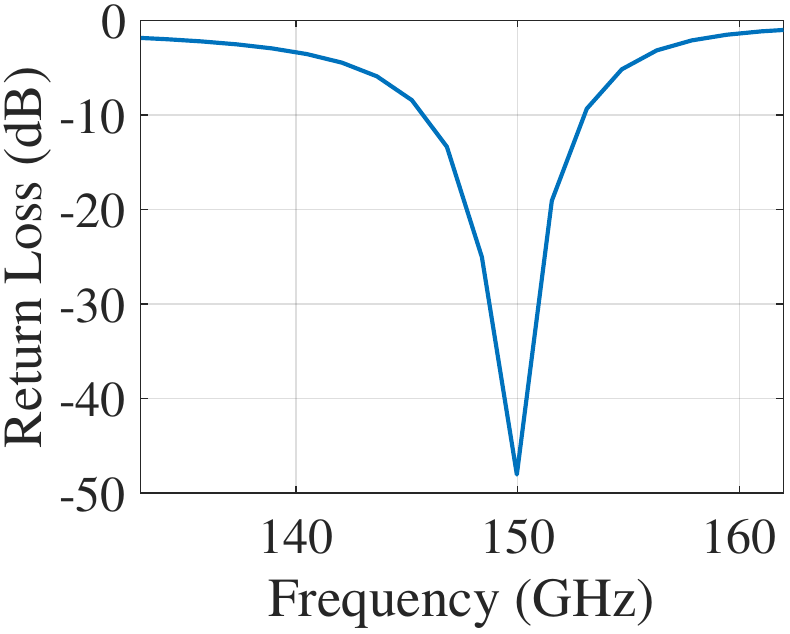}\label{fig:MPA-return-loss}} 
    \subfigure[]{\includegraphics[width=0.26\linewidth]{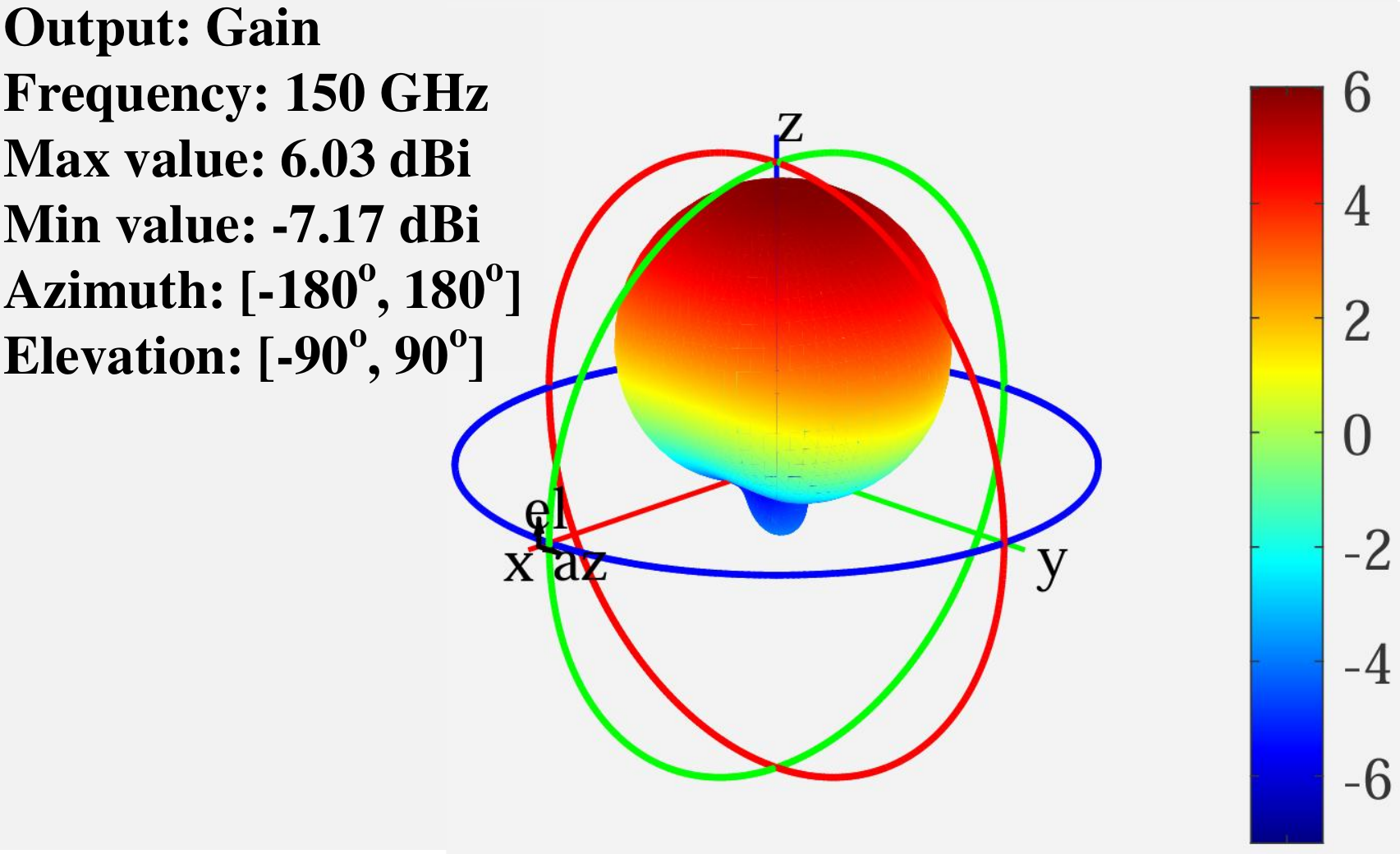}\label{fig:MPA-3d-pattern}}
    \subfigure[]{\includegraphics[width=0.18\linewidth]{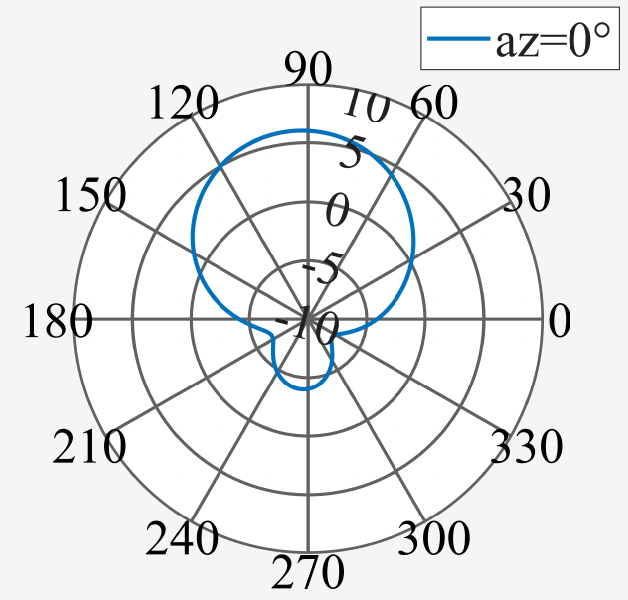}\label{fig:MPA-directivity}}
    \caption{Simulated (a) microstrip patch antenna using Foam as the substrate to resonate at a frequency of $150$ GHz in paint, (b) IP-MPA return loss variation with the frequency, (c) gain, and (d) the directivity (which depicts the gain in the polar coordinate system), implemented using MATLAB software.}
    \label{fig:Characteristics_of_MPA}
\end{figure*}

\begin{figure}[t]
    \centering
    \includegraphics[width=0.9\linewidth]{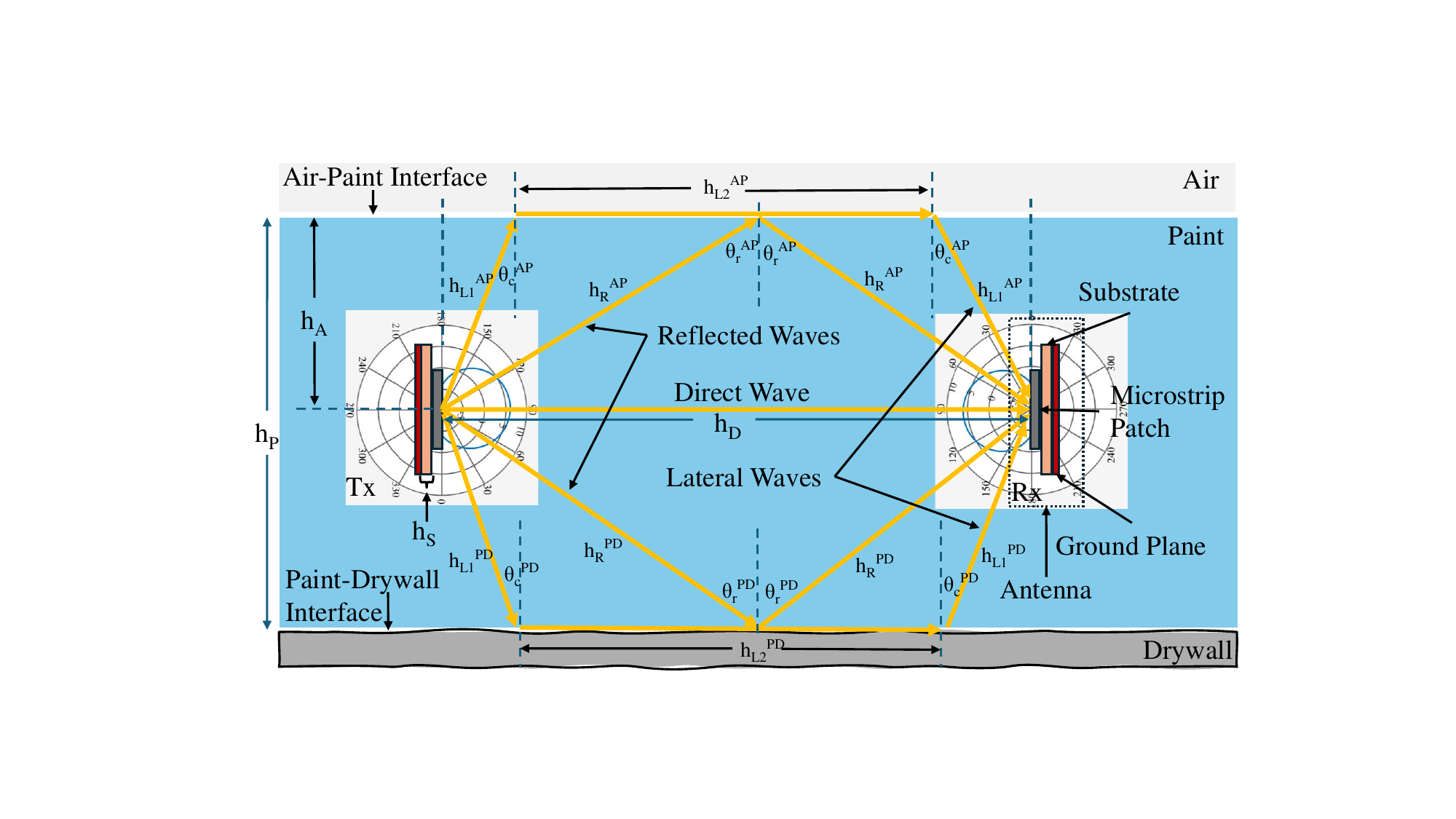}
    \caption{IoP: Embedded microstrip patch antenna multipath communication through paint.}
    \label{fig:MainFig}
\end{figure}

The IP-MPA is simulated using the Matlab Antenna Toolbox. The antenna design, return loss inside the paint, its gain, and directivity are shown in Figs.~\ref{fig:Characteristics_of_MPA}. The antenna dimensions in Table \ref{tab:MPA_dimensions} and Fig. \ref{fig:MPA-design} indicate that the size of the IP-MPA will be approximately a square millimetre, and a micron-sized antenna will be needed for higher frequencies. The return loss results (Fig.~\ref{fig:MPA-return-loss}) indicate that the IP-MPA will resonate at a frequency of $150$ GHz as expected. Moreover, the bandwidth of the IP-MPA, considering a $10$ dB return loss threshold, is  $7.07$ GHz ($146.03$ GHz - $153.1$ GHz) when embedded in the paint. Further, a maximum gain of $6.03$ dBi can be achieved at $150$ GHz (Figs.~\ref{fig:MPA-3d-pattern} and \ref{fig:MPA-directivity}) into the $95\degree$ direction. 

The IP-MPA performance will be evaluated based on an IoP channel model. First, we provide a brief overview of the multipath IoP channel model~\cite{Thakshila_IoP}, then we evaluate IP-MPA performance in Section~\ref{sec:Result_and_Discussio}. 

%\commentmcv{this analysis is backwards. It is using an over the air antenna and shows the change in resonant frequency when embedded in paint. However, our goal is to design an antenna that will operate at 150GHz when submerged in paint. This is not shown. What is the novelty here? This analysis basically suggests to change the operational frequency every time a different paint is used. This is not the goal. The goal is to be able to design embedded antennas tailored for the paint. This is not an IP-MPA design as claimed in the title or the abstract or the intro.}
%\commentBB{If I recall correctly, 150GHz was chosen because it was the highest frequency that the Matlab simulator could handle. The resonant frequency in paint is 8-9GHz less than in the antenna, so it would be better, as MCV says, if we could increase the ``antenna frequency'' (apologies if this terminology is incorrect\ldots) so that the desired resonant frequency (150GHz) is achieved. Perhaps TW could find a way to predict the antenna frequency for a desired resonant frequency (Eq. 4 seems to be a step in the right direction), but it is unclear whether frequencies higher than 150GHz can be realized by the Matlab simulator. One possibility would be to select 140GHz, say, as the desired resonant frequency, compute the corresponding antenna frequency, and rerun the simulations. However, there might not be time to rerun the experiments (paper deadline is 1 April), and it is likely to be more difficult to arrange now that TW is back in his teaching role at the University of Ruhuna. Comments welcome!!}

\section{Background: IoP Channel Model}
\label{sec:IoP_Channel_Model}

A comprehensive channel model for IoP is developed in~\cite{Thakshila_IoP} to capture sub-THz operation, accounting for unequal burial depths of the transceivers. Next, we provide an overview of this model. For details, we refer the reader to~\cite{Thakshila_IoP}. We assume that the technology is available to bury the transceivers at equal depths below the air-paint interface and focus on the effects of antenna orientation.
%\footnote{Arbitrary burial depths of transceiver-pairs are out of scope since our focus is on antenna orientations, and the reader is referred to~\cite{Thakshila_IoP}.}. 
The transceiver-pair architecture in paint is illustrated in Fig.~\ref{fig:MainFig}, where the embedded IP-MPAs interact with three mediums: air, paint, and drywall. These three mediums also create two interfaces: air-paint (A-P) and paint-drywall (P-D) interfaces. IoP communication can be established through five paths because properties such as the dielectric constant and wave propagation speed differ between media~\cite{Thakshila_IoP}. Thus, the dominant communication paths between two transceivers are (i) direct wave (DW), (ii) reflected wave from the A-P (RW-A) and (iii), the P-D (RW-D) interfaces, (iv) lateral wave through the A-P (LW-A) and (v), the P-D (LW-D) interfaces. Note that, the receiver IP-MPA can receive signals from all directions and the gain depends on the directivity. Based on the assumptions and the settings illustrated in Fig.~\ref{fig:MainFig}, we can express the path loss (dB) and the received power (dBm) for each dominant communication path as follows.

\subsection{Direct Wave}
\label{subsec:DW}

The path loss for the direct wave can be calculated by adding the spreading loss and absorption loss caused by the propagation in the paint medium \cite{Thakshila_IoP}, as follows:
\begin{equation}
\label{eq:Pathloss_direct}
    PL_D (dB)= 20 \log_{10}{\left(\frac{4\pi f_r h_D }{c_p}\right)} + 10\log_{10}{e^{h_D K_p(f_r)}},
\end{equation}
where $f_r$ is the resonant frequency of the IP-MPA, $h_D$ is the line of sight (LoS) distance between the transceivers, and $K_p(f_r)$ is the frequency-dependent absorption coefficient of paint, which can be found using (2) in \cite{Thakshila_IoP}. 

\begin{figure}
    \centering
    \includegraphics[width=0.9\linewidth]{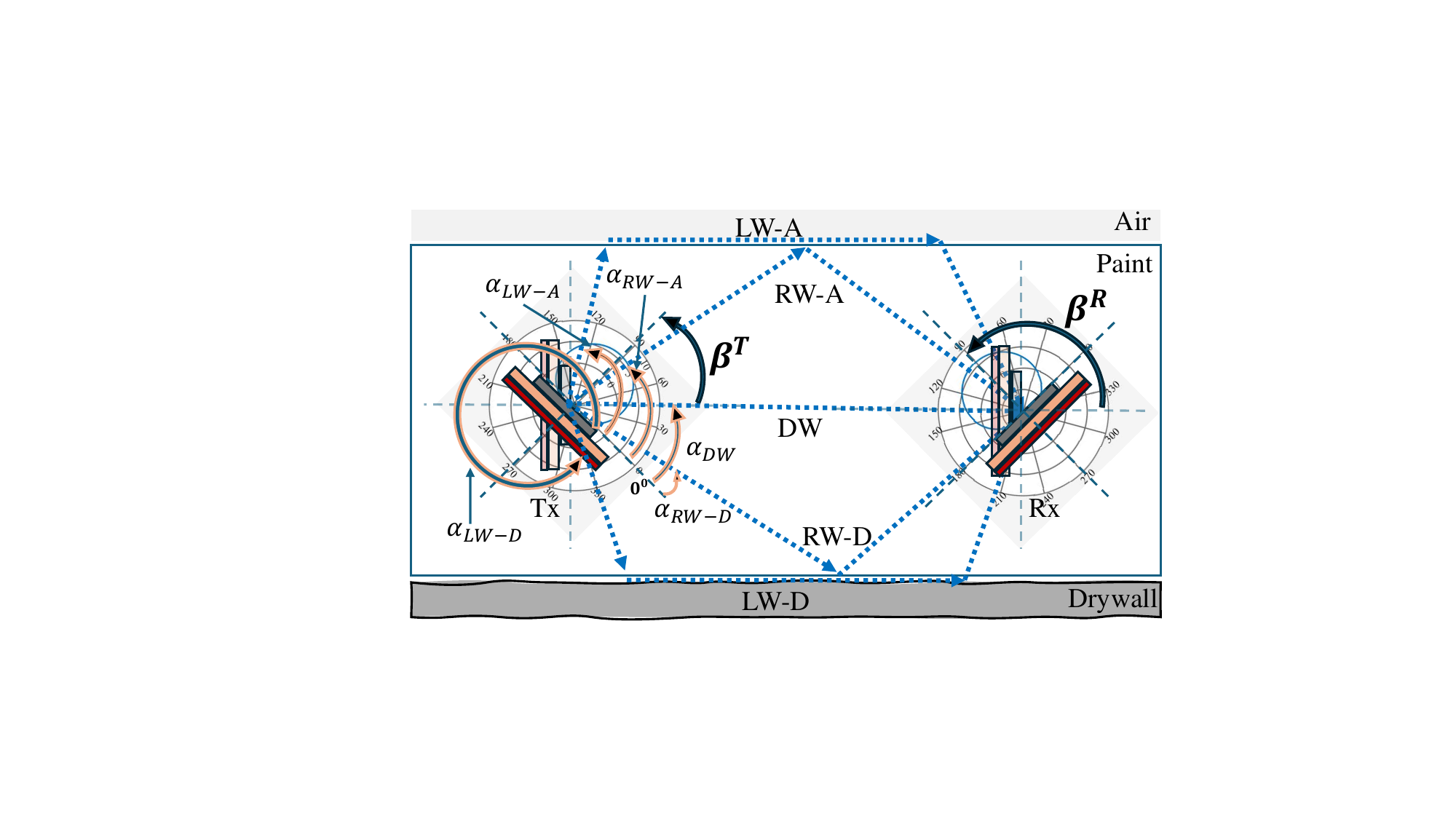}
    \caption{Variation in the boresight angle for the transceivers.}
    \label{fig:MPA_Tilting_by_beta}
\end{figure}
\begin{figure*}[t!]
    \centering
    \subfigure[LW-A]{\includegraphics[width=0.32\linewidth]{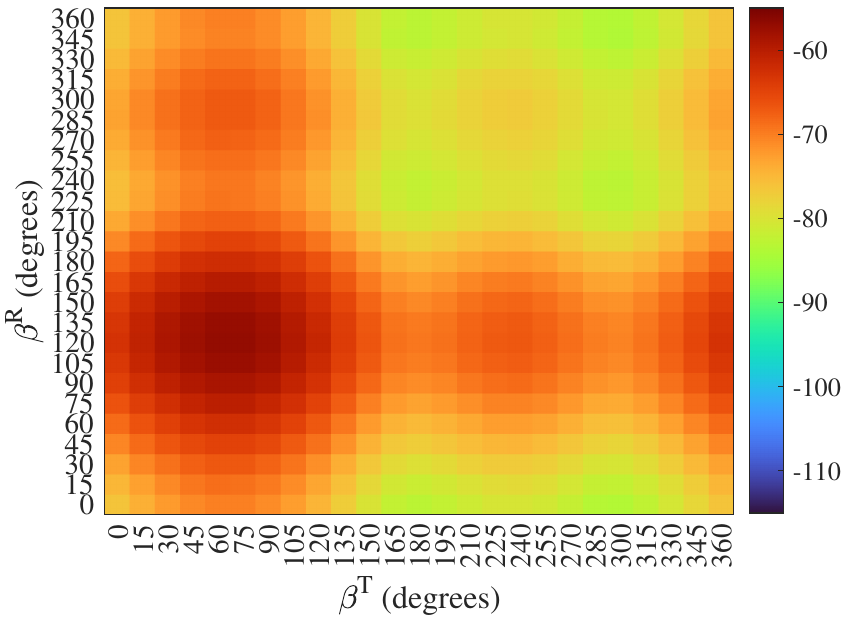}\label{subfig:heat_LW_A}} 
    \subfigure[RW-A]{\includegraphics[width=0.32\linewidth]{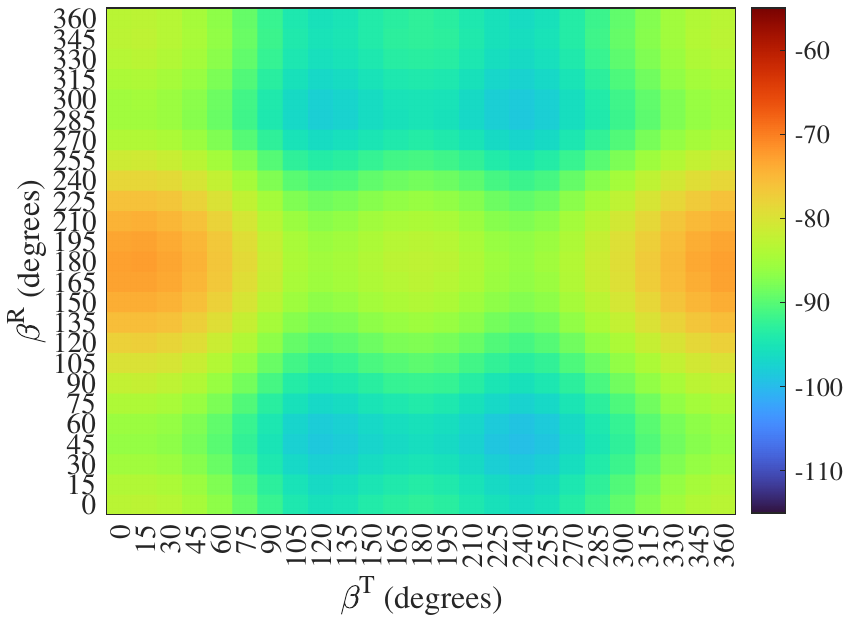}\label{subfig:heat_RW_A}} 
    \subfigure[DW]{\includegraphics[width=0.32\linewidth]{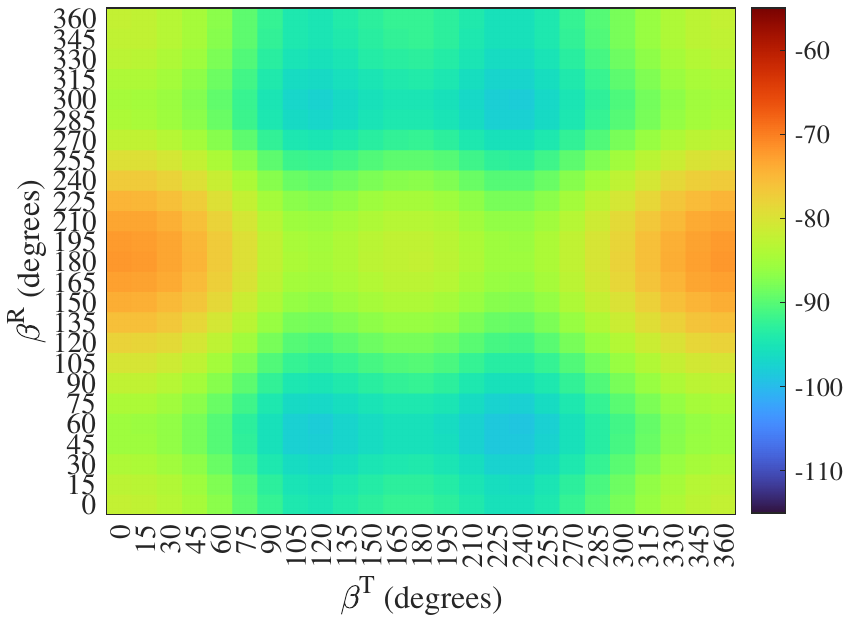}\label{subfig:heat_DW}}
    \subfigure[RW-D]{\includegraphics[width=0.32\linewidth]{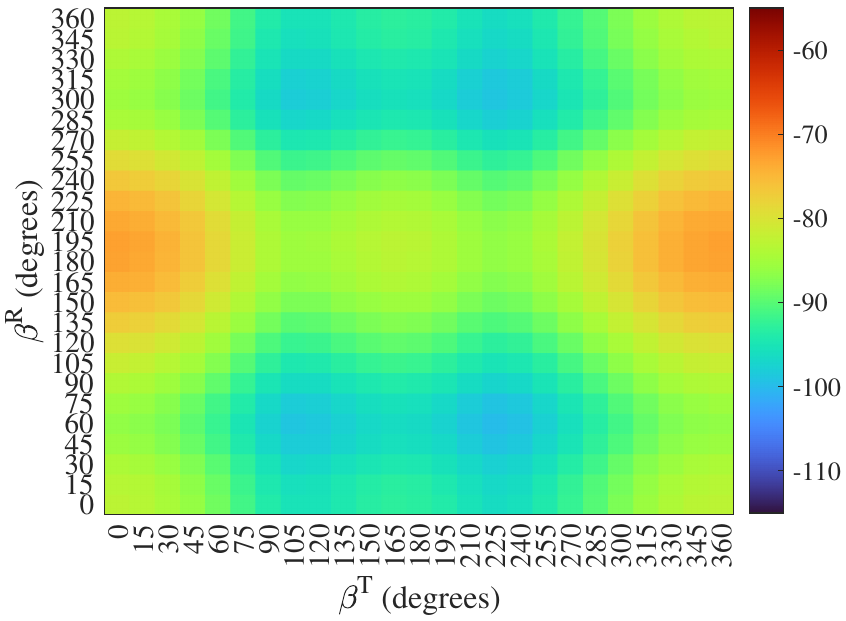}\label{subfig:heat_RW_D}}
    \subfigure[LW-D]{\includegraphics[width=0.32\linewidth]{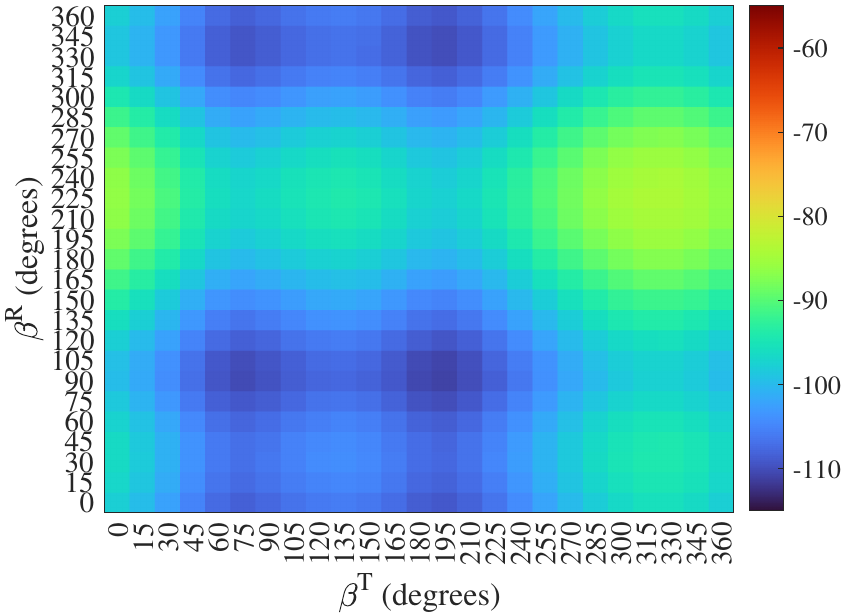}\label{subfig:heat_LW_D}}
    \subfigure[Total]{\includegraphics[width=0.32\linewidth]{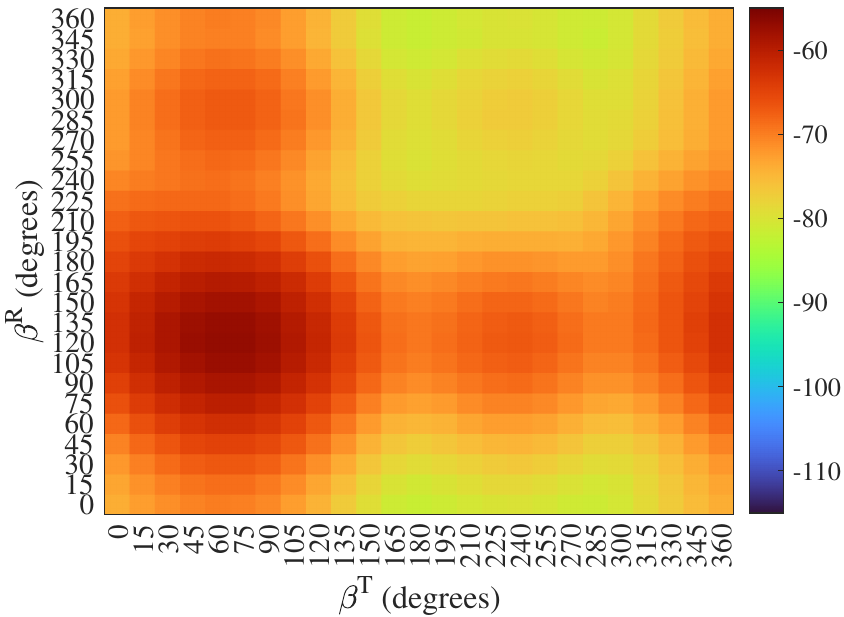}\label{subfig:heat_Total}}
    \caption{Total and multipath received power variation with $\beta^T$ and $\beta^R$ ($0\degree-360\degree$) when the MPA burial depth is $0.25$ mm.}
    \label{fig:Heat_Total_multi_RP_tilt_0_360}
\end{figure*}

\subsection{Reflected Wave}
\label{subsec:RW}

The refractive index of titanium dioxide white paint ($n_p = 2.13$) \cite{Thakshila_IoP} is greater than the refractive index of air ($n_a$ = $1$) and drywall ($n_{d} = 1.61$) \cite{refractive2008}. Thus, two reflected paths from the A-P and P-D interfaces are possible for the IoP communication. The angles $\theta_r^{AP}$ and $\theta_r^{PD}$ (see Fig.~\ref{fig:MainFig}) of the reflected paths can be determined using Snell's law for total internal reflection, which should exceed the critical angles $\theta_c^{AP}=\sin^{-1}(n_{a}/n_{p})$ ($\approx 28\degree$) and $\theta_c^{PD}=\sin^{-1}(n_{d}/n_p)$ ($\approx 49\degree$) corresponding to the A-P and P-D interfaces, respectively. Similar to the DW, the reflected waves RW-A and RW-D propagate within the paint medium. Therefore, when calculating the path loss, the spreading and absorption losses caused by the propagation of reflected waves (RW) in the paint medium are considered. Additionally, we also account for the surface roughness of the paint and drywall layers. Thus, considering all three components, the path loss for the RWs is \cite{Thakshila_IoP}:
\begin{equation}
\label{eq:reflectedPL}
\begin{split}
    PL^{i}_{R} (dB) = & 20 \log_{10}\left(\frac{8\pi f_r h^{i}_{R} }{c_p}\right) \\
    &+ 10\log_{10}{e^{2h^{i}_{R}K_p(f_r)}} - 10\log_{10}{R^{i}(f_r)},
\end{split}
\end{equation}
where $i$ represents the A-P or P-D interfaces, $h^{AP}_{R}=h_A/\cos(\theta_r^{AP})$, $h^{PD}_{R}=(h_P-h_A)/\cos(\theta_r^{PD})$, $h_P$ is the thickness of the paint layer and $h_A$ is the burial depth of the antennas. $R^{i}(f_r)$ is the frequency-dependent reflection coefficient, which can be found using (5) in \cite{Thakshila_IoP} considering paint for the A-P interface and drywall for the P-D interface as a rough surface.

\subsection{Lateral Wave}
\label{subsec:LW}

The lateral wave (LW) occurs when a wave propagates from a high refractive index medium to a lower refractive index medium and is incident at the critical angle~\cite{Vuran2019theoretical}. When a wave is released from the antenna embedded in a high refractive index medium and reaches the critical angle, it travels along the interface of the two mediums while releasing the energy of the electromagnetic wave back into the high refractive index medium at the critical angle direction. Since the paint medium's refractive index is higher than that of the air and drywall, two lateral paths propagating along the A-P and P-D interfaces are possible for IoP communication. However, unlike the DW and RWs, the LWs do not propagate entirely in the paint medium. After reaching the A-P (or P-D) interface, the wave propagates through the lower refractive index medium, which is air (or drywall). Thus, when calculating the path loss, we consider the spreading and absorption loss due to both mediums for each LW. Thus, the path loss is:
\begin{equation} 
\label{eq:lateral-approx}
\begin{split}
PL^{i}_{L} (dB) = &20 \log_{10}{\left(\frac{8\pi f_r h^{i}_{L1}}{c_p}  \right)}
+ 10\log_{10}{e^{2K_p(f_r) h^{i}_{L1}}}  \\
 &+ 20 \log_{10}{\left(\frac{4\pi f_r h^{i}_{L2}}{c_{i}}\right)} + 10\log_{10}{e^{K_{i}(f_r) h^{i}_{L2}}},
\end{split}
\end{equation}
where $i$ represents the A-P or P-D interfaces. When considering the LW-A, $h^{AP}_{L1} = h_A/\cos(\theta_c^{AP})$, $c_i = c ~(3 \times 10^8 ms^{-1})$ is the speed of the wave in vacuum, $h^{AP}_{L2} = h_D - 2h_A \tan(\theta_c^{AP})$, and $K_{AP}(f_r)$ is the frequency-dependent molecular absorption coefficient of air, which can be determined using (11) in \cite{Thakshila_IoP}. This calculation considers ten atmospheric gases that are commonly found in high concentrations and the data obtained from the HITRAN database~\cite{Hitran2016}. On the other hand, when considering the LW-D, $h^{PD}_{L1} = (h_P - h_A)/\cos(\theta_c^{PD})$, $c_{d}=c/n_d $ 
%{\bf**where is $c_{d}$?}\commentTW{The notation 'i' in the eq. \ref{eq:lateral-approx} represent A-P and P-D interfaces. For A-P interface $c_i=c$ and P-D interface $c_i = c_d$} 
is the speed of wave in drywall, $h^{PD}_{L2} = h_D - 2(h_P-h_A) \tan(\theta_c^{PD})$, and $K_{PD}(f_r)$ is the frequency-dependent absorption coefficient of drywall, which can be determined using (16) in \cite{Thakshila_IoP}.

Thus, the received power when utilizing each of these possible communication paths separately is \cite{josep2017terahertz}:
\begin{equation}
    \label{eq:Received_power}
    P_{R_{j}} (dBm)= P_t (dBm) +G_t(\alpha^{\beta^T}_j) +G_r(\alpha^{\beta^R}_j) - PL_j,
\end{equation}

where, $j$ stands for the DW, RW-A, RW-D, LW-A, and LW-D communication paths, $P_t$ is the transmit power, $PL_{j}$ represents the corresponding path loss of each path, and $G_t (\alpha^{\beta^T}_j)$ and $G_r (\alpha^{\beta^R}_j)$ are the transmitter and the receiver gains into the direction of multipath $\alpha_j$ (measured counter-clockwise from $0\degree$ directivity) corresponding to the boresight of the transmitter ($\beta^T$) and receiver ($\beta^R$) (see Fig.~\ref{fig:MPA_Tilting_by_beta}).  

Therefore, the total received power is \cite{Thakshila_IoP,dong2011channel}:
\begin{equation}
    \label{eq:Total_Received_power}
    P^{T}_R (dBm) = 10 \log_{10}{ \sum_{j}{10^{(P_{R_j}/10)}}}
\end{equation}

\section{Evaluation Results}
\label{sec:Result_and_Discussio}

In this section, we conduct a comprehensive numerical evaluation of the antenna based on multipath and total received power in relation to the boresight angles and burial depth of the transceivers. The gain in the direction of the multipath varies depending on the orientation of the MPA (Fig.~\ref{fig:MPA-directivity}). The thickness of the paint layer is $5$ mm, which is determined considering the substrate width ($931$ $\mu$m) and the length ($824$ $\mu$m). This allows the transceiver to be placed approximately $\pm 2$ mm from the middle of the paint layer in any orientation without protruding from the paint's surface. This functional paint layer is about twice as thick as conventional wall paint. 
%\commentmcv{This is a bit backwards. One might ask, why did we not determine the dimensions of the antenna based on a typical paint thickness and then decided on the frequency. It looks like 150GHz was chosen arbitrarily and now we are trying to fit the paint to that. It's too late to change it for this paper but something to consider going forward.}\commentTW{Noted, Professor}. 
However, unlike traditional paint, the proposed surface coating provides high-speed connectivity for all THz-enabled devices. The LoS distance between transceivers is fixed at $5$ cm, with the transmitter power considered to be $10$ mW ($10$ dBm)~\cite{josep2017terahertz}.

%%%%

\begin{table}[t!]
    \centering
    \addtolength{\tabcolsep}{-1pt} 
    \begin{tabular}{|l|c|c|c|c|c|}
    \hline
   % &  &  \multicolumn{2}{|c|}{Best Orientation}&  \multicolumn{2}{|c|}{Worst Orientation}\\
   % \hline
     & Burial & Maximum & ($\beta^T$, $\beta^R$)& Minimum& ($\beta^T$, $\beta^R$)\\
     & Depth & RP (dBm) & & RP (dBm)& \\
    \hline
         & $0.6$ & $-41.05$   & ($67\degree$,$123\degree$)& $-67.99$   & ($296\degree$,$352\degree$)\\
    LW-A & $2.5$ & $-56.88$   & ditto                     & $-83.83$   & ditto                     \\
         & $4.4$ & $-65.21$   & ditto                     & $-92.16$   & ditto                     \\
    \hline    
         & $0.6$  & $-72.14$  & ($6\degree$,$183\degree$)& $-99.09$    & ($235\degree$,$52\degree$)\\
    RW-A & $2.5$  & $-72.50$  & ($10\degree$,$179\degree$)& $-99.45$   & ($239\degree$,$48\degree$)\\
         & $4.4$  & $-73.51$  & ($15\degree$,$175\degree$)& $-100.46$   & ($244\degree$,$44\degree$)\\
    \hline
         & $0.6$  & $-72.03$   & ($5\degree$,$185\degree$)& $-98.98$   & ($234\degree$,$54\degree$)\\
    DW   & $2.5$  & ditto     & ditto                     & ditto   & ditto                     \\
         & $4.4$  & ditto      & ditto                    & ditto   & ditto                     \\
    \hline
         & $0.6$  & $-73.93$  & ($355\degree$,$195\degree$)& $-100.88$   & ($224\degree$,$64\degree$)\\
    RW-D & $2.5$  & $-72.75$  & ($359\degree$,$190\degree$)& $-99.70$   & ($228\degree$,$59\degree$)\\
         & $4.4$  & $-72.20$  & ($3\degree$,$186\degree$)& $-99.15$   & ($232\degree$,$55\degree$)\\
    \hline
         & $0.6$  & $-91.56$  & ($324\degree$,$225\degree$)& $-118.51$   & ($193\degree$,$94\degree$)\\
    LW-D & $2.5$  & $-84.11$  & ditto                      & $-111.06$   & ditto                     \\
         & $4.4$  & $-69.09$  & ditto                      & $-96.04$   & ditto                     \\
    \hline
               & $0.6$   & $-41.05$  & ($67\degree$,$123\degree$)  & $-67.89$   & ($295\degree$,$352\degree$)\\
        Total  & $2.5$   & $-56.85$  & ($66\degree$,$123\degree$)  & $-81.74$   & ($179\degree$,$351\degree$)\\
               & $4.4$   & $-64.96$  & ($63\degree$,$125\degree$)  & $-85.84$   & ($206\degree$,$347\degree$)\\
          \hline
    \end{tabular}
    \caption{\label{tab:RP_analysis_BD_Angle} The orientations of the transceivers ($\beta^T$, $\beta^R$) influence the multipath and the total, maximum and minimum RP, relating to the burial depth (in mm). (RP: Received Power).}
\end{table}

\subsection{Transceiver Boresight Analysis}
\label{subsec:Transceiver_Boresight_Analysis}

In this section, we analyze the multipath received power (RP) and total received power (TRP) variation based on arbitrary deployment orientations. To this end, we evaluate boresight angles of $[0\degree - 360\degree]$ for both the transmitter and receiver independently. The goal is to find best boresight angles with the highest TRP as well as the highest RP for each path. In Figs.~\ref{fig:Heat_Total_multi_RP_tilt_0_360}, RPs for each multipath (Figs.~\ref{subfig:heat_LW_A}-\ref{subfig:heat_LW_D}) as well as TRP (Fig.~\ref{subfig:heat_Total}) are shown. In the figures, the x and y axes show the boresight angle of the transmitter and receiver anntennas, respectively, and the received power is color coded for each angle pair. We consider the scenario where the transceivers are buried in the middle of the paint layer ($h_A = 2.5$ mm), and the transceivers (and their antennas) are oriented independently from $0\degree$ to $360\degree$. The maximum and minimum RP achieving orientations of the transceiver for each multipath and TRP for various burial depths are summarized in Table \ref{tab:RP_analysis_BD_Angle} with their corresponding boresight angles. 

It can be observed from Fig.~\ref{subfig:heat_LW_A} that LW-A has significantly less path loss compared to the other four paths, as most of the wave propagates in the lower absorbing medium of air~\cite{Thakshila_IoP}. LW-A RP is maximized ($\sim 57$ dBm) when the MPA orientation is ($\beta^T=67\degree$, $\beta^R=123\degree$), which is the propagation direction of the LW-A path. Additionally, the orientation angle range $(\beta^T \in (22.5\degree,127.5\degree) \cap \beta^R\in (67.5\degree,187.5\degree)) \cup (\beta^T \in (352.5\degree,360\degree) \cap \beta^R\in (82.5\degree,172.5\degree))$ shows relatively high ($ > -70$ dBm) RPs for IoP communication because the main lobe is directed along the communication path. However, the RP of LW-A is lower in $(\beta^T \in (157.5\degree,360\degree) \cap \beta^R\in (0\degree,37.5\degree)) \cup (\beta^T \in (157.5\degree,360\degree) \cap \beta^R\in (217.5\degree,360\degree))$, which includes the worst-performing orientation of the transceivers. Compared to other communication paths, the RP for LW-A is expected to be $15$ dB higher. 

The RW-A, DW, and RW-D paths (Figs.~\ref{subfig:heat_RW_A}--\ref{subfig:heat_RW_D}) yield similar RPs because they are close to each other, given the relatively long LoS distance ($5$cm) compared to the paint layer thickness ($5$ mm). This is the reason for higher RP when $(\beta^T \in (0\degree,52.5\degree) \cap \beta^R\in (152.5\degree,217.5\degree)) \cup (\beta^T \in (322.5\degree,360\degree) \cap \beta^R\in (152.5\degree,217.5\degree))$. Also, the antenna gain is minimal when the directivity is around $210\degree$ and $330\degree$ (see Fig. \ref{fig:Characteristics_of_MPA} (d)). At these angles, the RP decreases by $20-25$ dB for the three communication paths (RW-A, DW and RW-D) in the orientation ranges ($67.5\degree < \beta^T < 275.5\degree) \cap (0\degree < \beta^R < 97.5\degree)$ and ($67.5\degree < \beta^T < 275.5\degree) \cap (247.5\degree < \beta^R < 360\degree)$, compared to the corresponding maximum RPs, respectively. This is also the reason for the four RP minimums, which can be observed in the Figs. \ref{subfig:heat_RW_A}, \ref{subfig:heat_DW}, \ref{subfig:heat_RW_D} and also the Fig. \ref{subfig:heat_LW_D} for RP corresponding to the LW-D.   

Generally, the LW-D path has the lowest RP ($-111.06$ dBm), except when the transmitter and receiver boresights align with that path, where a $26.95$ dB higher RP can be expected. Thus, a higher LW-D path RP is expected when ($\beta^T \in (277.5\degree,360\degree) \cup \beta^T \in (0\degree,37.5\degree)) \cap \beta^R\in (172.5\degree,277.5\degree)$, but in all other regions its expected RP is significantly lower by $4-27.23$ dB compared to other communication paths.

In Fig.~\ref{subfig:heat_Total}, the contributions of each communication path to the total received power are shown. Clearly, LW-A is the dominant path. Moreover, direct and reflected paths add diversity to the feasible communication orientations of the IP-MPA. For instance, the total RP increases compared to the LW-A RP, especially when $\beta^R\in (7.5\degree,217.5\degree)$.

\subsection{Transceiver Burial Depth Analysis}
\label{subsec:Transceiver_Burial_Depth_Analysis}

Next, we analyze how RP varies with the transceiver burial depth based on the results in Section~\ref{subsec:Transceiver_Boresight_Analysis}. Table \ref{tab:RP_analysis_BD_Angle} summarizes the IP-MPA orientations that attain the maximum and minimum RP from each path and the multipath when transceivers are buried at $0.6$ mm, $2.5$ mm, $4.4$ mm, thereby placing the transceivers closer to the A-P, middle and closer P-D interfaces, respectively. The table indicates that the orientations of maximum and minimum RPs for RW-A and RW-D exhibit a dependency on depth, in contrast to the other paths, which are depth-independent. As expected, transceivers placed near the A-P interface provide a higher TRP of $15.8$ dB and $23.91$ dB compared to those placed in the middle and near the P-D interface, respectively. When the transceivers are positioned near the P-D interface, the consistent boresight rotation angles indicate that LW-A remains the dominant communication path, despite the increased communication distance. However, the RP for the LW-D path increases dramatically with burial depth.

\begin{figure}[t!]
    \centering
    \includegraphics[width=0.85\linewidth]{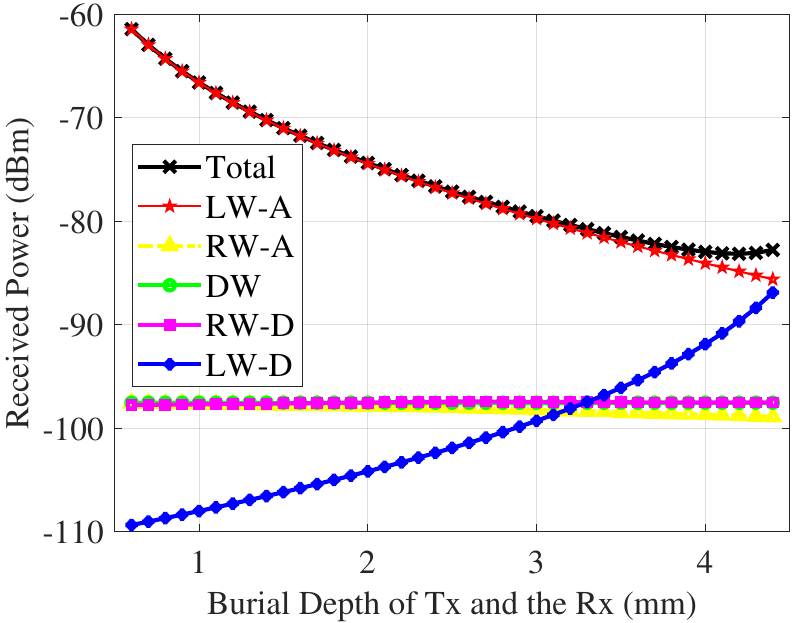}
    \caption{Variations in the best total received power with orientations of $\beta^T=66\degree$ and $\beta^R=123\degree$ compared to LW-A, RW-A, DW, RW-D, and LW-D paths with the same orientation angles.}  
    \label{fig:Total_and_multipath_best}
\end{figure}

Given the necessity for optimal antenna orientations in communication systems, we prioritize the maximum RP. Based on an evaluation of depth-dependent TRP, we determine that orientations of ($\beta^T=66\degree$, $\beta^R=123\degree$) represent the most advantageous configuration for the IoP at a burial depth of $2.5$ mm. Thus, Fig. \ref{fig:Total_and_multipath_best} illustrates the variation between the Total RP with the best orientation and each path RP with varying burial depths ($0.6$ mm - $4.4$ mm). Similar to the observation in Section~\ref{subsec:Transceiver_Boresight_Analysis}, LW-A is the dominant path with relatively high RP, averaging $-76.16$ dB for the same orientation as the Total path. The RP decreases with the burial depth from $-61.44$ dBm to $-85.61$ dBm due to the increasing communication distance and absorption caused by the paint medium. The TRP curve overlaps the RP curve corresponding to the dominant LW-A until a burial depth of $2$\,mm, and as it approaches the P-D interface, the difference increases to $2.85$\,dB. The reason for this behaviour is the exponential increase in the RP for the LW-D path, which can be observed in Fig. \ref{fig:Total_and_multipath_best} and Table \ref{tab:RP_analysis_BD_Angle} as the burial depth increases. Additionally, the RP corresponding to the LW-D path dominates other non-LW-A paths when the burial depth exceeds $3.3$ mm. Further, we observe that the RPs corresponding to LWs are approximately equal when the transceivers are placed near the P-D interface.

Finally, the RP associated with the DW remains constant at $-97.53$ dBm, as it is independent of the burial depth. As expected, the RP corresponding to RW-A is higher when transceivers are placed near the A-P interface, and with an increase in burial depth, RW-D becomes dominant. This behaviour is apparent in Table \ref{tab:RP_analysis_BD_Angle}, where the RPs corresponding to RW-A and RW-D increase and decrease, respectively, with increasing burial depth due to variations in propagation distance and absorption by the paint.

\section{Conclusion}
\label{sec:conclusion}

In this study, we develop a microstrip patch antenna embedded in a dielectric superstrate of paint with a resonant sub-THz frequency of $150$ GHz. We extend our channel model for the Internet of Paint (IoP) concept, in which devices are embedded seamlessly within a paint layer applied on a drywall to establish communication. This channel model predicts the total received power assuming the transmitter and receiver are buried at the same depth below the A-P interface. The variation of the total (multipath) received power is simulated for all boresight angles of the transceivers. The results show that ($\beta^T=66\degree$, $\beta^R=123\degree$) is the orientation of the transmitter/receiver pair that achieves the highest total received power. The LW-A is the most reliable single communication path for IoP, accounting for the majority of the total received power. Moreover, the LW-D path also demonstrates promising characteristics for IoP communication. Simulation results indicate that conventional walls covered by an IoP network of transceivers, combined with the proposed patch antennas, could enable pervasive high-speed wireless communication.

\section*{Acknowledgment}
{\label{5}}

This work is supported in part by Science Foundation Ireland (SFI) and the Department of Agriculture, Food and Marine on behalf of the Government of Ireland (Grant Number [16/RC/3835] - VistaMilk), and US National Science Foundation (NSF) ECCS-2030272, CNS-2212050, and CBET-2316960 grants.

\ifCLASSOPTIONcaptionsoff
  \newpage
\fi

\bibliographystyle{IEEEtran}

\bibliography{References}

\begin{thebibliography}{10}
\providecommand{\url}[1]{#1}
\csname url@samestyle\endcsname
\providecommand{\newblock}{\relax}
\providecommand{\bibinfo}[2]{#2}
\providecommand{\BIBentrySTDinterwordspacing}{\spaceskip=0pt\relax}
\providecommand{\BIBentryALTinterwordstretchfactor}{4}
\providecommand{\BIBentryALTinterwordspacing}{\spaceskip=\fontdimen2\font plus
\BIBentryALTinterwordstretchfactor\fontdimen3\font minus \fontdimen4\font\relax}
\providecommand{\BIBforeignlanguage}[2]{{%
\expandafter\ifx\csname l@#1\endcsname\relax
\typeout{** WARNING: IEEEtran.bst: No hyphenation pattern has been}%
\typeout{** loaded for the language `#1'. Using the pattern for}%
\typeout{** the default language instead.}%
\else
\language=\csname l@#1\endcsname
\fi
#2}}
\providecommand{\BIBdecl}{\relax}
\BIBdecl

\bibitem{Thakshila_IoP}
L.~T.~Wedage \emph{et~al.}, ``Internet of paint ({IoP}): Channel modeling and capacity analysis for terahertz electromagnetic nanonetworks embedded in paint,'' \emph{IEEE Journal on Selected Areas in Communications}, vol.~42, no.~8, pp. 2108--2121, 2024.

\bibitem{khan2020high}
M.~A.~K. Khan \emph{et~al.}, ``High-performance graphene patch antenna with superstrate cover for terahertz band application,'' \emph{Plasmonics}, vol.~15, no.~6, pp. 1719--1727, 2020.

\bibitem{younssi2013study}
M.~Younssi \emph{et~al.}, ``Study of a microstrip antenna with and without superstrate for terahertz frequency,'' \emph{International Journal of Innovation and Applied Studies}, vol.~2, no.~4, pp. 369--371, 2013.

\bibitem{bahl1982design}
I.~Bahl \emph{et~al.}, ``Design of microstrip antennas covered with a dielectric layer,'' \emph{IEEE Transactions on antennas and propagation}, vol.~30, no.~2, pp. 314--318, 1982.

\bibitem{Concrete2016analysis}
G.~Castorina \emph{et~al.}, ``Analysis and design of a concrete embedded antenna for wireless monitoring applications [antenna applications corner],'' \emph{IEEE Antennas and Propagation Magazine}, vol.~58, no.~6, pp. 76--93, 2016.

\bibitem{Wedage_IoP_challenges}
L.~Thakshila~Wedage, M.~C. Vuran, B.~Butler, C.~Argyropoulos, and S.~Balasubramaniam, ``Internet of paint ({IoP}): Design, challenges, applications, and future directions,'' \emph{IEEE Access}, vol.~13, pp. 31\,016--31\,023, 2025.

\bibitem{saidulu2013characteristics}
V.~Saidulu \emph{et~al.}, ``The characteristics of rectangular and square patch antennas with superstrate,'' \emph{International Journal of Engineering Sciences \& Emerging Technologies}, vol.~6, no.~3, pp. 298--307, 2013.

\bibitem{qasem2020simulation}
N.~Qasem and H.~M. Marhoon, ``Simulation and optimization of a tuneable rectangular microstrip patch antenna based on hybrid metal-graphene and fss superstrate for fifth-generation applications,'' \emph{TELKOMNIKA (Telecommunication Computing Electronics and Control)}, vol.~18, no.~4, pp. 1719--1730, 2020.

\bibitem{refractive2008}
C.~Jansen \emph{et~al.}, ``The impact of reflections from stratified building materials on the wave propagation in future indoor terahertz communication systems,'' \emph{IEEE Transactions on Antennas and Propagation}, vol.~56, no.~5, pp. 1413--1419, 2008.

\bibitem{Vuran2019theoretical}
A.~Salam, M.~C. Vuran, X.~Dong, C.~Argyropolous, and S.~Irmak, ``A theoretical model of underground dipole antennas for communications in internet of underground things,'' \emph{IEEE Transactions on Antennas and Propagation}, vol.~67, no.~6, pp. 3996--4009, 2019.

\bibitem{Hitran2016}
I.~Gordon \emph{et~al.}, ``\BIBforeignlanguage{en}{The {HITRAN2016} {Molecular Spectroscopic Database}},'' \emph{\BIBforeignlanguage{en}{Journal of Quantitative Spectroscopy and Radiative Transfer}}, vol. 203, pp. 3--69, Dec. 2017.

\bibitem{josep2017terahertz}
H.~Elayan \emph{et~al.}, ``Terahertz channel model and link budget analysis for intrabody nanoscale communication,'' \emph{IEEE transactions on nanobioscience}, vol.~16, no.~6, pp. 491--503, 2017.

\bibitem{dong2011channel}
X.~Dong and M.~C. Vuran, ``A channel model for wireless underground sensor networks using lateral waves,'' in \emph{2011 IEEE Global Telecommunications Conference-GLOBECOM 2011}.\hskip 1em plus 0.5em minus 0.4em\relax IEEE, 2011, pp. 1--6.

\end{thebibliography}

\end{document}